\documentclass[aps,prl,twocolumn,showpacs,superscriptaddress]{revtex4}
\usepackage{graphicx}
\begin{document}

\title{Topological Nature of the Phonon Hall Effect}

\author{Lifa~Zhang}
\email{zhanglifa@nus.edu.sg}
\affiliation{Department of Physics and Centre for Computational
Science and Engineering, National University of Singapore, Singapore
117542, Republic of Singapore }
\author{Jie Ren}
\email{renjie@nus.edu.sg}
\affiliation{NUS Graduate School for Integrative Sciences
and Engineering, Singapore 117456, Republic of Singapore}
 \affiliation{Department of
Physics and Centre for Computational Science and Engineering,
National University of Singapore, Singapore 117542, Republic of
Singapore }
\author{Jian-Sheng~Wang}
\email{phywjs@nus.edu.sg}
\affiliation{Department of Physics and Centre for Computational
Science and Engineering, National University of Singapore, Singapore
117542, Republic of Singapore }
\author{Baowen~Li}
\email{phylibw@nus.edu.sg}
\affiliation{NUS Graduate School for Integrative Sciences
and Engineering, Singapore 117456, Republic of Singapore}
 \affiliation{Department of
Physics and Centre for Computational Science and Engineering,
National University of Singapore, Singapore 117542, Republic of
Singapore }

\date{02 Aug 2010, Revised 25 Nov 2010}

\begin{abstract}
We provide a topological understanding of the phonon Hall
effect in dielectrics with Raman spin-phonon coupling. A general
expression for phonon Hall conductivity is obtained in terms of the
Berry curvature of band structures. We find a nonmonotonic behavior
of phonon Hall conductivity as a function of the magnetic field.
Moreover, we observe a phase transition in the phonon Hall effect, which
corresponds to the sudden change of band topology, characterized by
the altering of integer Chern numbers. This can be explained by
touching and splitting of phonon bands.
\end{abstract}


\maketitle

Recent years have witnessed a rapid development of an emerging field
-- phononics, the science and technology of controlling heat flow
and processing information with phonons \cite{Wang2008}. Indeed, in
parallel with electronics, various functional thermal devices such as
thermal diode \cite{diode}, thermal transistor \cite{transistor},
thermal logic gates \cite{logic} and thermal memory \cite{memory},
etc., have been proposed to manipulate and control phonons, the
carrier of heat energy and information. However, different from
electrons, phonons as neutral quasiparticles, cannot directly
couple  to the magnetic field through the Lorentz force. Therefore,
it is a surprise that Strohm, Rikken, and Wyder observed the
phonon Hall effect (PHE) - the appearance of a temperature difference in the direction perpendicular to both the applied magnetic field and the heat current flowing through an ionic paramagnetic dielectric sample \cite{strohm05}. It was confirmed later by Inyushkin and Taldenkov \cite{inyushkin07}. Since then, several theoretical
explanations have been proposed \cite{shengl06,kagan08,wang09} to
understand this novel phenomenon.

For electronic transport properties in various
quantum, spin, or anomalous Hall effects \cite{TKNN, QSHE, AHE},
topological Berry phase has been successfully used to understand the
underlying mechanism \cite{xiao10}. Such an elegant connection
between mathematics and physics provides a broad and deep
understanding of basic material properties. However, because of the
very different nature of electrons and phonons, a topological
picture related to the PHE is not straightforward and obvious, and
therefore, is still lacking.

In this Letter, we explore the topology of phonon bands in a
two-dimensional honeycomb lattice with Raman type spin-phonon
interaction. A general expression for phonon Hall conductivity in
terms  of Berry curvature is derived. The phonon Hall effect is not
quantized, although the Chern numbers are quantized to integers. We
find that there exists a phase transition associated with the PHE, due
to the discontinuous jump of Chern numbers.

We start with a Hamiltonian for an ionic crystal lattice in a uniform external magnetic field \cite{holz72}, which reads in a compact form as
\begin{eqnarray}
H & = &\frac{1}{2} (p-{ \tilde A }
u)^T (p-{ \tilde A }u) + \frac{1}{2} u^T K u  \qquad\nonumber  \\
& = &\frac{1}{2} p^T p + \frac{1}{2} u^T (K-{ \tilde A }^{2}) u + u^T\! { \tilde A }\,p.
\label{eq-model}
\end{eqnarray}
Here, $u$ is a column vector of
displacements from lattice equilibrium positions for all the
degrees of freedom, multiplied by the square root of mass, $p$ is
the conjugate momentum vector, and $K$ is the force constant matrix.
The superscript $T$ stands for the matrix transpose. ${ \tilde A }$ is
an antisymmetric real matrix,  which is block diagonal with elements
 $ \Lambda=
\left(\begin{array}{rr} 0 & h  \\
-h & 0  \\
\end{array}\right)
$ (in two dimensions), where $h$ is proportional to the magnitude of the applied magnetic field, and has the dimension of frequency. For simplicity, we will call $h$ the magnetic field later. The on-site term, $u^T{ \tilde A }p$, can be
interpreted as the Raman (or spin-phonon) interaction \cite{supp}.
The Hamiltonian (1) is positive definite.

By applying Bloch's theorem, we can describe the system by the
polarization vector $x =(\mu, \epsilon)^T$, where $\mu $ and
$\epsilon $ are associated with the momenta and coordinates,
respectively. The equation of motion can be expressed as
\begin{equation}\label{eq-eom}
i\frac{\partial }
{{\partial t}}x  = H_{\rm eff} x, \;\;  H_{\rm eff}= i\left( \begin{array}{cc} -A & -D \\
I & -A \end{array} \right),
\end{equation}
where $D({\bf k}) = - A^2+\sum_{l'} K_{l,l'} e^{i({\bf R}_{l'} -
{\bf R}_{l})\cdot {\bf k}}$ is the dynamic matrix as a function of
wave vector ${\bf k}$; $K_{l,l'}$ is the submatrix between unit cell
$l$ and $l'$ in the full spring constant matrix $K$; ${\bf R}_l$ is
the real-space lattice vector; $A$ is block diagonal with elements
 $\Lambda$, and $I$ is an identity matrix. Here, $D, A, K_{l,l'}$, and $I$ are all  $4\times4$ matrices for the two-dimensional honeycomb lattice.
The eigenvalue problem of the equation of motion (\ref{eq-eom}) reads:
\begin{equation}
H_{\rm eff}\, x_\sigma = \omega_\sigma\, x_\sigma, \label{eq-eigen}
\end{equation}
where $ x_{\sigma} = (\mu_{\sigma}, \epsilon_{\sigma})^T$ is the
right eigenvector of the $\sigma$-th branch and $\omega_{\sigma}$ is
the corresponding eigenfrequency. Because of the non-Hermitian nature of
$H_{\rm eff}$, the left eigenvector is different, and is given by
${\tilde x_{\sigma}}^T = (\tilde \mu_{\sigma}, \tilde
\epsilon_{\sigma}) = (\epsilon^\dagger_{\sigma}, -
\mu^\dagger_{\sigma})/( - 2i\omega _\sigma )$.  The orthonormal
condition is $\epsilon_{\sigma}^{\dag}\epsilon_{\sigma'}+
\frac{i}{\omega_{\sigma}}\epsilon_{\sigma}^{\dag}A\epsilon_{\sigma'}=\delta_{\sigma,\sigma'}$
\cite{wang09}.

By taking into account only positive eigenfrequency modes, displacement and
momentum operators can be written in the second quantization form.
From the definition of energy current density ${\bf J} =
\frac{1}{2V} \sum_{l,l'} ({\bf R}_l -{\bf R}_{l'}) u^T_{l} K_{l,l'}
\dot{u}_{l'}$ \cite{hardy63,shengl06,kagan08}, the current density
vector can be expressed as
\begin{equation}
{\bf J} = {\bf J_1}(a^\dagger a)+{\bf J_2}(a^\dagger a^\dagger, a
a).
\end{equation}
Here, ${\bf J_1} =\frac{\hbar}{4V} \sum\limits_{k,k'} {
\frac{\omega_k+\omega_{k'}} {\sqrt{\omega_k\omega_{k'}}}
\epsilon_k^\dagger \frac{\partial D({\bf k})}{\partial {\bf k}}
\epsilon_{k'}\, a_k^\dagger a_{k'} e^{i(\omega _k  - \omega _{k'}
)t}}\delta_{{\bf k},{\bf k}'}$, and $ {\bf J_2} =\frac{\hbar }
{{4V}}\sum\limits_{k,k'} {\sqrt {\frac{{\omega _{k'} }} {{\omega _k
}}} } \,(\epsilon _k^\dag \frac{{\partial D({\bf k})}} {\partial
{\bf k}}\epsilon _{k'}^ * a_k^\dag  a_{k'}^\dag e^{i(\omega _k  +
\omega _{k'} )t}  + \,\epsilon _k^T \frac{{\partial D^
*  ({\bf k})}} {\partial {\bf k}}\epsilon _{k'}^{} a_k^{} a_{k'}^{} e^{ - i(\omega _k  + \omega _{k'} )t} )\delta _{{\bf k}, - {\bf k}'}$, where
$k=({\bf k},\sigma)$ considers both the wavevector and the phonon branch.
It should be noted that the $a^\dagger a^\dagger$ and $aa$ terms also
contribute to the off-diagonal elements of thermal conductivity
tensor, although they have no contribution to the average heat flux.
The diagonal term $\epsilon_k^\dagger \frac{\partial D({\bf
k})}{\partial {\bf k}} \epsilon_{k}$ in ${\bf J_1}$ corresponds to
$\omega_\sigma \frac{\partial \omega_\sigma}{ \partial {\bf k}}$.
Only the off-diagonal terms in ${\bf J_1}$ and ${\bf J_2}$
contribute to the Hall conductivity, which can be regarded as the
contribution from anomalous velocities similar to the one in the
intrinsic anomalous Hall effect \cite{AHE}. Using the Green-Kubo
formula $\kappa_{xy} = \frac{V}{\hbar T} \int_0^{\beta\hbar}
d\lambda \int_0^\infty dt\, \bigl\langle J^x(-i\lambda) J^y(t)
\bigr\rangle_{\rm eq}$ \cite{mahan00}, one can obtain phonon Hall
conductivity as \cite{supp}:
\begin{eqnarray}
\kappa_{xy} &=& \frac{\hbar }
{{8V T}}\sum\limits_{\sigma\neq\sigma '} {f(\omega _\sigma ) (\omega _\sigma  + \omega _{\sigma '})^2 } \times \qquad\nonumber  \\
&& \frac{i} {{4 \omega _\sigma  \omega _{\sigma '} }}\frac{\epsilon
_\sigma ^\dag \frac{{\partial D}} {{\partial k_x }}\epsilon _{\sigma
'} \epsilon _{\sigma '}^\dag \frac{{\partial D}} {{\partial k_y
}}\epsilon _\sigma - (k_x \leftrightarrow k_y ) } {(\omega _\sigma
- \omega _{\sigma '} )^2 }, \label{eq-kxy}
\end{eqnarray}
where $f(\omega _\sigma ) = (e^{\hbar \omega _\sigma/(k_B T)} -
1)^{-1}$ is the Bose distribution function, $V$ is the total
volume of the sample, and the phonon branch index $\sigma$ here
includes both the positive and negative eigenvalues without
restrictions. It can be proved that the phonon Hall conductivity
$\kappa_{xy}$ satisfies the Onsager reciprocal relations \cite{supp}.

\begin{figure}[t]
\includegraphics[width=1.0\columnwidth, height= 2.0 in ]{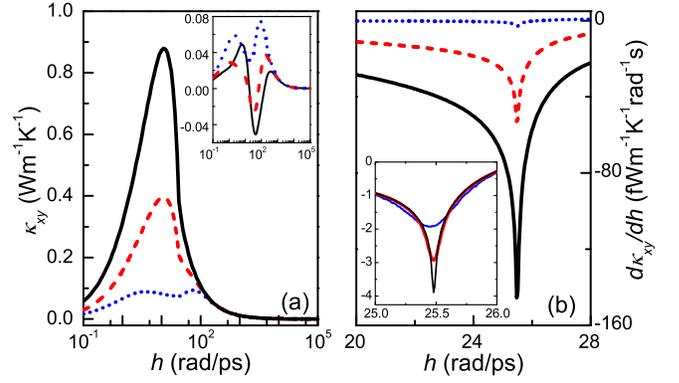}
\caption{ \label{fig1cond} (color online) (a) Phonon Hall
conductivity vs magnetic field for different temperatures. The
dotted, dashed, and solid lines correspond to $T=50, 100$, and $300$
K, respectively. The inset shows $h$-dependence of $\kappa_{xy}$ at
low temperatures: $T=10$ (solid line), $20$ (dashed line), and $40$ K
(dotted line). (b) $d\kappa_{xy}/dh$ as a function of $h$ at different
temperatures: $T=50$ (dotted line), $100$ (dashed line), and $300$ K
(solid line); here $N_L=400$. The inset in (b) shows the $h$-dependence
of $d\kappa_{xy}/dh$ for different size $N_L$ at $T=50$ K, around
$h\approx 25.5$ rad/ps; from top to bottom, $N_L=80, 320$, and
$1280$, respectively.}
\end{figure}

In Fig.~\ref{fig1cond} we show the phonon Hall conductivity of
honeycomb lattices calculated from Eq.~(\ref{eq-kxy}). The
parameters used in our numerical calculations are the same as in
Ref.~\cite{wang09}. The coupling matrix between two sites is
configured such that the longitudinal spring constant is $K_L =
0.144\,$eV/(u\AA$^2$) and the transverse one $K_T$ is 4 times
smaller.  The unit cell lattice vectors are $(a,0)$ and $(a/2,
a\sqrt{3}/2)$ with $a=1\,$\AA.

It is found that when $h$ is small, $\kappa_{xy}$ is proportional to
$h$ \cite{supp}, while the dependence becomes nonlinear when  $h$ is
large. As $h$ is further increased, $\kappa_{xy}$ increases before
it reaches a maximum at  certain value of $h$. Then $\kappa_{xy}$
decreases and goes to zero at very large $h$. This can be understood
as follows: numerical calculation shows that $\omega_\sigma \approx
\alpha h$, which can also be obtained from the equation $\bigl[
(-i\omega_\sigma + A)^2 + D\bigr] \epsilon_\sigma = 0$ \cite{supp},
thus we can obtain approximately $\kappa_{xy} \sim
h^2/(e^{\beta\hbar\alpha h}-1)$ from Eq.~(\ref{eq-kxy}). In the weak
magnetic field limit $\kappa_{xy}\propto h$, while in the strong
field limit, $\kappa_{xy}\rightarrow 0$. The on-site term $\tilde{A}^2$ in
the Hamiltonian (\ref{eq-model}) increases with $h$ quadratically so
as to blockade the phonon transport, which competes with the
spin-phonon interaction. Therefore, as $h$ increases, $\kappa_{xy}$
first increases, then decreases and tends to zero at last. At low
temperatures, $\kappa_{xy}$ oscillates around zero with the
variation of $h$, as shown in the inset in Fig.~\ref{fig1cond}(a).

There is a subtle singularity near $h\simeq25$ rad/ps in
Fig.~\ref{fig1cond}(a); we thus plot the first derivative of
$\kappa_{xy}$ with respect to $h$ at different temperatures in
Fig.~\ref{fig1cond}(b). It shows that, at the relatively high
temperatures, the first derivative of phonon Hall conductivity has a
minimum at the magnetic field $h_c\simeq25.4778$ rad/ps for the
finite-size sample $N_L=400$ (the sample has $N=N_L^2$ unit cells).
The first derivative $d\kappa_{xy}/dh$ at the point $h_c$ diverges
when the system size increases to infinity. The inset in
Fig.~\ref{fig1cond}(b) shows the finite-size effect. At the point
$h_c$, the second derivative $d^2\kappa_{xy}/dh^2$ is
discontinuous. Therefore, $h_c$ is a critical point for the PHE, across
which a phase transition occurs. At low temperatures, the divergence
of $d\kappa_{xy}/dh$ is not so evident as that at high temperatures.
However, if the sample size becomes larger, the discontinuity of
$d^2\kappa_{xy}/dh^2$ is more obvious, as illustrated in
Fig.~\ref{fig1cond}(b). For different temperatures, the phase
transition occurs at exactly the same critical value $h_c$, which
strongly suggests that the phase transition of the PHE is related to the
topology of the phonon band structure.

In the following, we would like to connect the PHE with the Berry phase
to examine the underlying topological mechanism. As is wellknown,
the band structure of crystals provides a natural platform to
investigate the geometric phase effect. Since the wave-vector
dependence of the polarization vectors is inherent to the Hall
problems, the Berry phase effects are intuitively expected for the PHE
in the momentum space. Following Berry's approach~\cite{xiao10}, we
set $x(t) = e^{i\gamma _\sigma  (t) - i\int_0^t {dt'\omega _\sigma({\bf
k}(t'))}} x _\sigma  ({\bf k}(t)) $, and then insert it into
Eq.~(\ref{eq-eom}). The Berry phase is obtained as $\gamma _\sigma =
\oint\limits \mathbf{A}^{\sigma}_{\mathbf{k}} \cdot d {\bf k}$, with
$ \mathbf{A}^{\sigma}_{\mathbf{k}}=i \tilde x _\sigma ^T
\frac{\partial x _\sigma} {{\partial {\bf k}}},$ and the Berry
curvature emerges as
\begin{equation}\label{eq-brycv}
\Omega _{k_x k_y }^\sigma =\frac{\partial } {{\partial
{k_x}}}\mathbf{A}^{\sigma}_{k_y}-\frac{\partial } {{\partial
{k_y}}}\mathbf{A}^{\sigma}_{k_x} =\sum\limits_{\sigma ',\sigma ' \ne
\sigma } {\Omega _{k_x k_y }^{\sigma \sigma '} },
\end{equation}
where,
\begin{equation}\label{eq-Omga}
\Omega _{k_x k_y }^{\sigma \sigma '}=\frac{i} {{4 \omega _\sigma
\omega _{\sigma '} }}\frac{\epsilon _\sigma ^\dag \frac{{\partial
D}} {{\partial k_x }}\epsilon _{\sigma '} \epsilon _{\sigma '}^\dag
\frac{{\partial D}} {{\partial k_y }}\epsilon _\sigma - \epsilon
_\sigma ^\dag \frac{{\partial D}} {{\partial k_y }}\epsilon _{\sigma
'} \epsilon _{\sigma '}^\dag  \frac{{\partial D}} {{\partial k_x
}}\epsilon _\sigma } {(\omega _\sigma   - \omega _{\sigma '} )^2 }
\end{equation}
is the contribution to the Berry curvature of the band $\sigma$ from a different band $\sigma '$. The associated topological Chern number is obtained through integrating the Berry curvature over the first Brillouin zone as
\begin{equation}\label{eq-cherni}
C^\sigma   = \frac{1} {{2\pi }}\int_{{\rm BZ}} {dk_x dk_y \Omega
_{k_x k_y }^\sigma  }= \frac{{2\pi }} {{L^2}}\sum\limits_{\bf k}
{\Omega _{k_x k_y }^\sigma  },
\end{equation}
where, $L$ is the length of the sample. The phonon Hall conductivity formula, Eq.~(\ref{eq-kxy}), is
recasted into
\begin{equation}
\kappa_{xy} = \frac{\hbar } {{8V T}}\sum\limits_{{\bf
k},\sigma\neq\sigma '} {f(\omega _\sigma ) (\omega _\sigma  + \omega
_{\sigma '})^2 }\Omega _{k_x k_y }^{\sigma \sigma '} .
\label{eq-main}
\end{equation}
Here $V=L^2a$. The term $(\omega _\sigma  + \omega _{\sigma '})^2$ relating to the phonon energy is an analog of the electrical charge term $e^2$ in
the electron Hall effect, thus the phonon Hall conductivity
Eq.~(\ref{eq-main}) is similar to but different from the electron
case because the phonon energy term can not be moved out from the
summation. Although the formula is derived from the phonon transport
in the crystal-lattice system, we note that the thermal Hall
conductivity for the magnon Hall effect \cite{katsura10} can also be
cast into the form of Eq.~(\ref{eq-main}) with a different
expression for the Berry curvature. Therefore, the Hall conductivity
formula can be universally applicable to the thermal Hall effect in
phonon and magnon systems without restriction for special lattice
structures.
\begin{figure}[t]
\includegraphics[width=1.00\columnwidth]{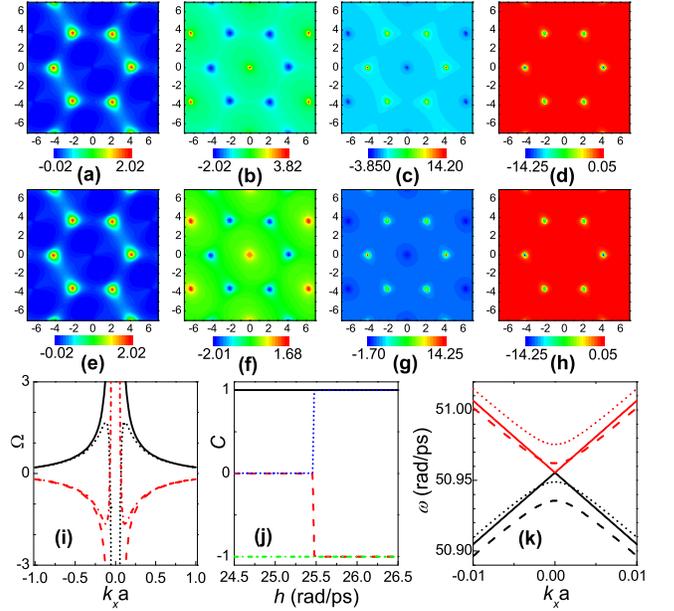}%
\caption{ \label{fig2toplg} (color online) (a)-(d) The contour map
of Berry curvatures for bands $1-4$ at $h_{c-}=h_c-10^{-2}$ rad/ps;
(e)-(h) The contour map of Berry curvatures for bands $1-4$ at
$h_{c+}=h_c+10^{-2}$ rad/ps. For (a)-(h), the horizontal and
vertical axes correspond to wave vector $k_x$ and $k_y$,
respectively. (i) $\Omega$ at different magnetic fields. The solid
and dashed lines correspond to $\Omega^2$ and $\Omega^3$ at
$h_{c-}$  respectively, while dotted and dash-dotted lines correspond to those at $h_{c+}$. (j) Chern numbers of four bands: $C^1$
(solid line), $C^2$ (dashed line), $C^3$ (dotted line), and $C^4$ (dash-dotted line).
(k) The dispersion relation of band $2$ and $3$ at different
magnetic fields in the vicinity of $h_c$. The dashed, solid and
dotted lines correspond to the bands at $h_{c-}$, $h_{c}$, and $h_{c+}$,  respectively. The lower three and upper three correspond to bands 2 and 3,
respectively. $k_y=0$ in (i) and (k). }
\end{figure}

Without the Raman spin-phonon interaction, namely, $h=0$, $\Omega
_{k_x k_y }^{\sigma \sigma '}$ is zero everywhere and the phonon
Hall conductivity vanishes. When a magnetic field is applied, the
Berry curvature is nonzero, and consequently, the PHE appears. It is
found that if the system exhibits symmetry satisfying $ SDS^{ - 1} =
D,$ $SAS^{ - 1} = - A$ (e.g., mirror reflection symmetry), the
phonon Hall conductivity is zero \cite{wang09,supp}. This symmetry principle can also
be applied to the topological property of the phonon bands: we find
that $\Omega _{k_x k_y }^{\sigma \sigma '}=0$ provided that such
symmetry exists, such as in the square lattice system. Whereas if
such symmetry is broken for the dynamic matrix, the system can
possess nontrivial Berry curvatures. In the system with the PHE, if
the magnetic field changes, the Berry curvatures are quite different.
However, we find that the associated topological Chern numbers
remain constant integers with occasional jumps when $h$ is varied.
Therefore, the Chern numbers given by Eq.~(\ref{eq-cherni}) are
topological invariant, which indeed illustrates the nontrivial
topology of the phonon band structures. Although the Chern numbers
are quantized to integers, the phonon Hall conductivity is not, due
to the extra term $f(\omega _\sigma ) (\omega _\sigma  + \omega
_{\sigma '})^2 $. Thus, the analogy to the quantum Hall effect is
incomplete.

In the vicinity of the critical magnetic field $h_c$, we find that
the phase transition is indeed related to the abrupt change of the
topology of band structures. The Berry curvatures for different
bands near the critical magnetic field are illustrated in
Fig.~\ref{fig2toplg}(a-h). We find that with an infinitesimal change
of magnetic field around $h_c$, the Berry curvatures around the $\Gamma$ (${\bf k}=0$) point of bands $2$ and $3$ are quite different, whereas those
of band $1$ and $4$ remain unchanged. To illustrate the change of
the Berry curvatures clearly, we plot the cross section of the Berry curvatures along the
$k_x$ direction for bands $2$ and $3$ in
Fig.~\ref{fig2toplg}(i), which shows explicitly that the Berry
curvatures change dramatically above and below the critical magnetic
field $h_c$. Below the critical point, the Berry curvature for band
$2$ in the vicinity of $\Gamma$  point contributes Berry phase
$2\pi$ ($-2\pi$ for band $3$), which cancels that from $K$,and $K'$
points, so that the Chern number is zero for bands $2$ and $3$, as
indicated in Fig.~\ref{fig2toplg}(j). However, above the critical
point, the sum of Berry curvature at $\Gamma$ point is zero, and
only the monopole at $K$,and $K'$ points contributes to Berry phase
($-2\pi$ for band $2$ and $2\pi$ for band $3$). Therefore, the Chern
numbers jump from $0$ to $\pm 1$, as shown in
Fig.~\ref{fig2toplg}(j). This jump indicates that the topology of
the two bands suddenly changes at the critical magnetic field, which
is responsible for the phase transition. From a calculation on the
kagome lattice, which has been used to model many real materials
\cite{kagome}, we also find qualitatively similar phase transitions
due to the sudden change of topology, where the phonon Hall
conductivity has three singularities of divergent first derivatives
corresponding to three jumps of the Chern numbers.

To further investigate the mechanism of the abrupt change of the
phonon band topology, we study the dispersion relation near the
critical magnetic field. From Fig.~\ref{fig2toplg}(k), we can see
that band $2$ and $3$ are going to touch with each other at the $\Gamma$
point if the magnetic field increases to $h_c$; at the critical
magnetic field, the degeneracy occurs and the two bands possess the
cone shape; above the critical point $h_c$, the two bands split up.
Therefore, the difference between the two bands decreases below and
increases above the critical point $h_c$. The property of the
dispersion relation in the vicinity of the critical magnetic field
directly affects the Berry curvature of the corresponding bands.

In summary, we have studied the PHE from a topological point of
view. By looking at the phases of the polarization vectors of both the
displacements and conjugate momenta as a function of the wave vector, a
Berry curvature can be defined uniquely for each band. This Berry
curvature can be used to calculate the phonon Hall conductivity. Because of  the nature of phonons, the phonon Hall conductivity, which is not
directly proportional to the Chern number, is not quantized.
However, the quantization effect, in the sense of discontinuous
jumps in Chern numbers, manifests itself in the phonon Hall
conductivity as a singularity of the first derivative with respect
to the magnetic field.

The topological approach for phonon Hall conductivity proposed here
is general and can be applied to the real materials in low
temperatures where the thermal transport is ballistic. It can also
be applied to the magnon Hall effect discovered
recently \cite{katsura10}. Phase transition in the PHE, explained
from topological nature and dispersion relations, can also be
generalized to study the phase transition in other Hall effects
and/or nonequilibrium transport.  In line with recently reported
Berry-phase-induced heat pumping \cite{Ren10} and the
Berry-phase contribution of molecular vibrational instability
\cite{Lv10}, we hope our present results do invigorate the studies
aimed at uncovering intriguing Berry phase effects and topological
properties in phonon transport, which will enrich further the
discipline of phononics.

L.Z. thanks Bijay Kumar Agarwalla and Jie Chen for fruitful
discussions. This project is supported in part by Grants
No. R-144-000-257-112 and No. R-144-000-222-646 of NUS.

\vskip 5cm

\begin{widetext}
\setcounter{figure}{0}
\setcounter{equation}{0}
\setcounter{section}{0}

\renewcommand\thefigure{S\arabic{figure}}
\renewcommand\theequation{S\arabic{equation}}
\section{Supplementary information for ``Topological Nature of Phonon Hall Effect"}

\vskip 1.5cm
\subsubsection{ABSTRACT}
\emph{In this supplementary material, we discuss the origination of the
Hamiltonian [Eq. (1) in the main text] in the first section; then we
present the detailed derivation of the general formula of phonon
Hall conductivity in terms of Berry curvature, in which we also give
the explicit expression for the dynamic matrix and give the proof
for the symmetry principle. Finally, we discuss the numerical
calculation for Chern numbers.}
\vskip 1.5cm

\subsection{\label{seccond} DISCUSSION ON THE HAMILTONIAN}
In the presence of a magnetic field, according to
Ref.~\cite{holz72}, the kinetic energy of each site in ionic crystal
lattices without free charges is expressed as:
\begin{equation}
T_\alpha=\frac{1}{2} m_\alpha |{\dot{\bf r}}_\alpha|^2=\frac{1}{2
m_\alpha} |{\bf p}_\alpha\sqrt{m_\alpha}-q_\alpha {\bf A}_\alpha|^2,
\label{eq-skinetic}
\end{equation}
where, ${\bf r}_\alpha={\bf R}_\alpha+{\bf
u}_\alpha/\sqrt{m_\alpha}$, ${\bf R}_\alpha$ is the equilibrium
coordinate of the ion at site $\alpha$, and ${\bf u}_\alpha$ denotes
the displacement multiplied by the square root of the ion mass
$m_\alpha$. ${\bf p}_\alpha$ is the corresponding momentum divided
by the square root of mass $m_\alpha$. $q_\alpha$ is the ionic
charge at site $\alpha$. ${\bf A_\alpha}$ denotes the
electromagnetic vector potential, which, using the Lorenz gauge
condition, can be related to the ionic displacement as \cite{holz72}
\begin{equation}
{\bf A}_\alpha=\frac{1}{2}{ \bf B}\times {\bf u}_\alpha/\sqrt{m_\alpha}.
\end{equation}
Thus, Eq.~(\ref{eq-skinetic}) is recasted as:
\begin{equation}
T_\alpha=\frac{1}{2} |{\bf p}_\alpha- \frac{q_\alpha}{2 m_\alpha}{ \bf B}\times {\bf u}_\alpha|^2.
\end{equation}
If the magnetic field with magnitude $B$ is applied along $z$
direction and we only consider the two-dimensional ($x$ and $y$
direction) motion of the system, then the kinetic energy of ion
$\alpha$ can be expressed (it is straightforward to generalize to
high dimentions) as:
\begin{equation}
T_\alpha=\frac{1}{2} (p_\alpha- \Lambda_\alpha
u_\alpha)^T (p_\alpha- \Lambda_\alpha u_\alpha),
\end{equation}
where $p_\alpha=(p_{\alpha x},  p_{\alpha y})^T$,
$u_\alpha=(u_{\alpha x},  u_{\alpha y})^T$, and $ \Lambda_{\alpha}=
\left(\begin{array}{cc} 0 & h_\alpha  \\
-h_\alpha & 0  \\
\end{array}\right)
$, where $h_\alpha=-q_{\alpha}B/(2 m_{\alpha})$. Note that
there are both positive and negative ions in one unit cell. For a
general ionic paramagnetic dielectric, mostly, the mass of the
positive ion is larger than that of the negative one. For instance,
in the experimental sample ${\rm Tb_3 Ga_5 O_{12}}$, the ratio
$m(+q)/m(-q)$ is about $4.3$ in one unit cell. Therefore the
negative ions will dominate in the contribution to $h_{\alpha}$,
which makes $h_{\alpha}$ have the same sign as that of the applied
magnetic field $B$. Under the mean-field approximation, we can set
$h_\alpha=h$, which is site-independent and is proportional to the
magnitude of the applied magnetic field.

Combining the kinetic energy with the harmonic inter-potential energy,
we can write the whole Hamiltonian as
\begin{equation}
H = \frac{1}{2} (p-{ \tilde A }
u)^T (p-{ \tilde A }u) + \frac{1}{2} u^T K u, \\
\end{equation}
where  ${ \tilde A }$ is an antisymmetric real matrix with
block-diagonal elements $\Lambda_\alpha$.  $u$ and $p$ are column
vectors denoting displacements and momenta respectively, for all the
degrees of freedom. $K$ indicates the force constant matrix.
Finally, after the rearrangement, we have
\begin{equation}
H = \frac{1}{2} p^T p + \frac{1}{2} u^T (K-{ \tilde A }^{2}) u + u^T\! { \tilde A }\,p,
\label{eq-Sham}
\end{equation}
which is exactly the second row of Eq.~(1) in the text.

The Hamiltonian Eq. (\ref{eq-Sham}) [Eq.(1) in the main text] is
essentially the same as that used in
Ref.~\cite{shengl06,kagan08,wang09,zhang09} resulting from the
phenomenological Raman interaction. The only difference is the term
proportional to $\tilde{A}^2$ which makes the above Hamiltonian
positive definite. The Raman interaction, proposed to study
spin-phonon interactions (SPI) based on quantum theory and
fundamental symmetries \cite{old-sp-papers,sp-book,ioselevich95},
can be expressed as
\begin{equation}
 H_I=g {\bf s} \cdot ({\bf u} \times {\bf p}).
\end{equation}
Here, $g$ denotes a positive coupling constant, and ${\bf s}$ is the
isospin for the lowest quasidoublet. In the presence of a magnetic
field ${\bf B}$, each site has a magnetization ${\bf M}$. For
isotropic SPI, the isospin ${\bf s}$ is parallel to ${\bf M}$, and
the ensemble average of the isospin is proportional to the
magnetization, which can be expressed as $\langle{\bf s}\rangle$ =
$c {\bf M}$ with $c$ the proportionality coefficient
(Ref.~\cite{shengl06,kagan08,wang09,zhang09}). In the mean-field
approximation, the Raman type SPI reduces to
\begin{equation}
 H_I={\bf h} \cdot ({\bf u} \times {\bf p}),
\end{equation}
where ${\bf h}=gc{\bf M}$, and ${\bf M}$ is proportional to the
magnetic field ${\bf B}$. If the magnetic field is applied along the
$z$ direction, then the SPI can be written as
\begin{equation}
 H_I=u^T\,\tilde A\, p.
\end{equation}
By treating the phonon system under harmonic approximation, the
total Hamiltonian for the whole lattice can be written as
(Ref.~\cite{shengl06,kagan08,wang09})
\begin{equation}
H = \frac{1}{2} p^T p + \frac{1}{2} u^T K u + u^T\! { \tilde A }\,p.
\label{eq-sRamam}
\end{equation}
Note that this Hamiltonian Eq. (\ref{eq-sRamam}) is not positive
definite. In Ref.\cite{wang09}, the authors added an arbitrary
onsite potential in order to make the Hamiltonian positive definite.
However, in the calculation of phonon Hall effect for the
four-terminal junctions, such non-positive-definite Hamiltonian does
not cause any problem because the thermal junctions will stabilize
the system \cite{zhang09}.

From the first physical picture of spin-phonon interaction in ionic
crystal lattice with an applied magnetic field
(Eq.~\ref{eq-kinetic}$\sim$\ref{eq-Sham}), the additional term
proportional to ${\tilde A}^2$ emerges naturally to make the
Hamiltonian positive definite. Therefore, in this work we choose the
positive definite Hamiltonian Eq.~(\ref{eq-Sham}) [Eq. (1) in the
main text].

\subsection{PHONON HALL CONDUCTIVITY FROM GREEN-KUBO FORMULA}

The Hamiltonian Eq.~(\ref{eq-Sham}) is quadratic in $u$ and $p$, and
we can write the equation of motion as
\begin{eqnarray}\label{eq-seom1}
\dot p &=&  - (K - \tilde{A}^2 )u - \tilde{A}p, \\
\dot u &=& p - \tilde{A}u.
\label{eq-seom2}
\end{eqnarray}
The equation of motion  for the coordinate is,
\begin{equation}
\ddot u + 2\tilde{A}\dot u + \tilde{A}^2 u + (K - \tilde{A}^2 )u = 0.
\end{equation}
Since the lattice is periodic, we can apply the Bloch's theorem
$u_l=\epsilon e^{i ({\bf R}_{l} \cdot {\bf k}- \omega t)} $. The
polarization vector $\epsilon$ satisfies
\begin{equation}
\bigl[ (-i\omega + A)^2 + D\bigr] \epsilon = 0,
\label{eq-sdisper}
\end{equation}
where $D({\bf k}) = - A^2+\sum_{l'} K_{l,l'} e^{i({\bf R}_{l'} -
{\bf R}_{l})\cdot {\bf k}}$ denotes the dynamic matrix and $A$ is
block diagonal with elements $\Lambda$. $D, K_{l,l'},$ and $A$ are
all $n d\times n d$ matrices, where $n$ is the number of particles
in one unit cell and $d$ is the dimension of the motion.

To calculate the dynamic matrix $D({\bf k})$, we give an example for
the two-dimensional honeycomb lattice, where $n=2, d=2$. We only consider
the nearest neighbor interaction. The spring constant matrix along $x$ direction is
\begin{equation}
K_x=\left(\begin{array}{cc} K_L & 0  \\
0 & K_T  \\
\end{array}\right).
\end{equation}
$K_L = 0.144\,$eV/(u\AA$^2$) is the longitudinal spring constant and
the transverse one $K_T$ is 4 times smaller.  The unit cell lattice
vectors are $(a,0)$ and $(a/2, a\sqrt{3}/2)$ with $a=1\,$\AA.

To obtain the explicit formula for the dynamic matrix, we first
define a rotation operator in two dimensions as:
$$
U(\theta ) = \left( {\begin{array}{*{20}c}
   {\cos \theta } & { - \sin \theta }  \\
   {\sin \theta } & {\cos \theta }  \\
 \end{array} } \right).
$$
The three kinds of spring-constant matrices between two atoms are
$K_{01}  = U(\pi/2)K_x U( - \pi/2 )$, $K_{02}  = U(\pi /6)K_x U( -
\pi /6)$, $K_{03}  = U( - \pi /6)K_x U(\pi /6)$, which are
$2\times2$ matrices. Then we can obtain the on-site spring-constant
matrix and the four spring-constant matrices between the unit cell
and its four nearest neighbors as:
\begin{equation}
  K_0  = \left( {\begin{array}{*{20}c}
   {K_{01}  + K_{02}  + K_{03} } & { - K_{02} }  \\
   { - K_{02} } & {K_{01}  + K_{02}  + K_{03} }  \\
 \end{array} } \right),
\end{equation}
 \begin{equation}
  K_1  = \left( {\begin{array}{*{20}c}
   0 & 0  \\
   { - K_{03} } & 0 \end{array} } \right),\quad
  K_2  = \left( {\begin{array}{*{20}c}
   0 & 0  \\
   { - K_{01} } & 0  \\  \end{array} } \right),\quad
  K_3  = \left( {\begin{array}{*{20}c}
   0 & { - K_{03} }  \\
   0 & 0  \\ \end{array} } \right),\quad
   K_4  = \left( {\begin{array}{*{20}c}
   0 & { - K_{01} }  \\
   0 & 0  \\ \end{array} } \right), \\
 \end{equation}
which are $4\times4$ matrices.
Finally we can obtain the $4\times4$ dynamic matrix $D({\bf k})$ as
\begin{equation}\label{sDexplict}
D({\bf k}) =-A^2+ K_0  + K_1 e^{ik_x }  + K_2 e^{i(k_x /2 + \sqrt 3 k_y /2)}  + K_3 e^{ - ik_x }  + K_4 e^{ - i(k_x /2 + \sqrt 3 k_y /2)},
\end{equation}
where, $A^2=-h^2\cdot I$, and $I$ denotes the $4\times4$ identity
matrix.

Equation~(\ref{eq-sdisper}) can be written in a form as a standard
eigen-problem given in Eq.~(3) in the main text, if we rewrite the
equations of motion. Using Bloch theorem, Eqs.~(\ref{eq-seom1}) and
(\ref{eq-seom2}) can be recasted as:
\begin{equation}
i\frac{\partial }
{{\partial t}}x  = H_{\rm eff} x, \;\;\;\;\;\;\;\;\;\;  H_{\rm eff}= i\left( \begin{array}{cc} -A & -D \\
I & -A \end{array} \right).
\end{equation}
$x =(\mu, \epsilon)^T$ is the polarization vector , where column
vectors $\mu $ and $\epsilon $ are associated with the momenta and
coordinates respectively. Therefore, the right eigenvector and left
eigenvector satisfy:
\begin{equation}
H_{\rm eff} x_\sigma= \omega_\sigma x_\sigma, \;\;\;\;\;\;\;\;\;\;
\tilde{x}_\sigma^T H_{\rm eff} = \omega_\sigma \tilde{x}_\sigma^T.
\end{equation}
where  the right eigenvector $x_\sigma =(\mu_\sigma,
\epsilon_\sigma)^T$, the left eigenvector ${\tilde x_{\sigma}}^T =
(\epsilon^\dagger_{\sigma}, - \mu^\dagger_{\sigma})/( - 2i\omega
_\sigma )$, and $\sigma$ indicates the branch index. Because the
effective Hamiltonian $H_{\rm eff}$ is not hermitian, the
orthonormal condition then holds between the left and right
eigenvectors. The eigenmodes can be normalized as ${\tilde
x_{\sigma}}^T x_\sigma=1$, which is equivalent to \cite{wang09}
\begin{equation}
\epsilon_{\sigma}^\dagger \epsilon_{\sigma} +
\frac{i}{\omega_\sigma} \epsilon_{\sigma}^\dagger\! A
\epsilon_{\sigma}  = 1. \label{eq-snorm}
\end{equation}
To solve the eigensystem, we require the following relations:
\begin{equation}\label{eq-sew1}
\epsilon_{{-\bf k},-\sigma}^* = \epsilon_{{\bf k},\sigma}; \;
\omega_{{-\bf k},-\sigma} = -\omega_{{\bf k},\sigma}.
\end{equation}

In the following, we use $k=({\bf k},\sigma)$ to specify both the
wavevector and the phonon branch. By taking into account only
positive eigen-modes ($\omega>0$), displacement and momentum
operators are taken in the second quantization form:
\begin{eqnarray}
u_l &=& \sum_{k} \epsilon_k e^{i ( {\bf R}_l \cdot {\bf k}-\omega_k t)} \sqrt{\frac{\hbar}{2\omega_k N}}\; a_k + {\rm h.c.},\\
p_l &=& \sum_{k} \mu_k e^{i ({\bf R}_l \cdot {\bf k} -\omega_k t ) } \sqrt{\frac{\hbar}{2\omega_k N}}\; a_k + {\rm h.c.},
\end{eqnarray}
where $\sigma>0$, $a_k$ is the annihilation operator, and h.c.\
stands for hermitian conjugate.  The momentum and displacement
polarization vectors are related through $\mu_k = -i\omega_k
\epsilon_k + A\epsilon_k$.  We can verify that the canonical
commutation relations are satisfied: $[u_l, p_{l'}^T] = i\hbar
\delta_{l,l'} I$, and $H= \sum_{k} \hbar \omega_k (a_k^\dagger a_k +
1/2)$.

The energy current density is defined as \cite{hardy63}:
\begin{equation}
{\bf J}= \frac{1}{2V} \sum_{l,l'} ({\bf R}_l\! -\! {\bf R}_{l'})
u^T_{l} K_{l,l'} \dot{u}_{l'},
\end{equation}
where $V$ is the total volume of $N$ unit cells.  The current
density vector can be expressed in terms of the
creation/annihilation operators as
\begin{equation}
\begin{array}{ll}
{\bf J} = {\bf J}_1(a^\dagger a)+{\bf J}_2(a^\dagger a^\dagger, a a);\\
{\bf J}_1 =\frac{\hbar}{4V}
\sum_{k,k'} \left( \sqrt{\frac{\omega_k}{\omega_{k'}}} + \!
\sqrt{\frac{\omega_{k'}}{\omega_k}} \right)
\epsilon_k^\dagger \frac{\partial D({\bf k})}{\partial {\bf k}} \epsilon_{k'}\,a_k^\dagger a_{k'} \delta_{{\bf k},{\bf k}'} e^{i(\omega _k  - \omega _{k'} )t}
;\\
{\bf J}_2 =\frac{\hbar }
{{4V}}\sum\limits_{k,k'} {\sqrt {\frac{{\omega _{k'} }}
{{\omega _k }}} } \,(\epsilon _k^\dag  \frac{{\partial D({\bf k})}}
{\partial {\bf k}}\epsilon _{k'}^ *  a_k^\dag  a_{k'}^\dag  e^{i(\omega _k  + \omega _{k'} )t}  + \,\epsilon _k^T \frac{{\partial D^ *  ({\bf k})}}
{\partial {\bf k}}\epsilon _{k'}^{} a_k^{} a_{k'}^{} e^{ - i(\omega _k  + \omega _{k'} )t} )\delta _{{\bf k}, - {\bf k}'}.
\end{array}
\end{equation}
We note that the $a^\dagger a^\dagger$ and $aa$ terms also
contribute to the off-diagonal elements of the thermal conductivity
tensor, although they have no contribution to the average energy
current. Based on the expression of heat current, the phonon Hall
conductivity can be obtained through the Green-Kubo formula
\cite{mahan00}:
\begin{equation}
\kappa_{xy} = \frac{V}{\hbar T} \int_0^{\hbar/(k_B T)}\!\!\!\! d\lambda
\int_0^\infty\! dt\, \bigl\langle J^x(-i\lambda) J^y(t) \bigr\rangle_{\rm eq},
\label{eq-sGK}
\end{equation}
where the average is taken over the equilibrium ensemble with
Hamiltonian $H$. 
Substituting the expression ${\bf J}$ into Eq.~(\ref{eq-sGK}), the
phonon Hall conductivity is obtained as
\begin{equation}
\begin{array}{ll}
\kappa_{xy}= \kappa_{xy}^{(1)}+ \kappa_{xy}^{(2)};\\
\kappa_{xy}^{(1)}=\frac{V}{\hbar T} \int_0^{\hbar/(k_B T)}\,d\lambda \int_0^\infty\! dt\, \bigl\langle J_1^x(-i\lambda) J_1^y(t) \bigr\rangle_{\rm eq}
;\\
\kappa_{xy}^{(2)}=\frac{V}{\hbar  T} \int_0^{\hbar/(k_B T)}\, d\lambda \int_0^\infty\! dt\, \bigl\langle J_2^x(-i\lambda) J_2^y(t) \bigr\rangle_{\rm eq}
.
\end{array}
\end{equation}
Note that the averages of the cross terms $\bigl\langle
J_1^x(-i\lambda) J_2^y(t) \bigr\rangle_{\rm eq}$ and $\bigl\langle
J_2^x(-i\lambda) J_1^y(t) \bigr\rangle_{\rm eq}$ are zero.

First we calculate the term $\kappa_{ab}^{(1)}$. Combining the result
\begin{equation}
\langle a_i^\dagger a_j a_k^\dagger a_l \rangle_{\rm eq} =
f_i f_k \delta_{ij}\delta_{kl} + f_i (f_j+1)\delta_{il} \delta_{jk},
\end{equation}
where $f_i = (e^{\beta\hbar \omega_i} - 1)^{-1}$ is the Bose distribution
function, with the result
\[
\sum\limits_{\bf k} {\epsilon _\sigma ^\dag  \frac{{\partial D}}
{{\partial k_{\alpha}  }}\epsilon _\sigma  }  = -2i \sum\limits_{\bf
k} { \omega _\sigma {\tilde x_{\sigma}}^T  \frac{{\partial H_{\rm
eff}}} {{\partial k_{\alpha}  }} x _\sigma  }=-2i \sum\limits_{\bf
k} {\omega _\sigma  \frac{{\partial \omega _\sigma  }} {{\partial
k_{\alpha}  }} = 0},  \;\;\;\;\;\;\;\;(\alpha=x, y)
\]
which is obtained by differentiating the Eq.~(S20) (Eq.~(3) in the
main text) and $\tilde x_{\sigma}^T x _\sigma =1 $, we obtain
\begin{equation}
\kappa_{xy}^{(1)} = \frac{\hbar } {{16V  T}}\sum\limits_{{\bf
k},\sigma  > 0,\sigma ' > 0} {[f(\omega_\sigma ) - f(\omega_\sigma
')](\omega_\sigma + \omega_\sigma ')^2 \frac{i} {{\omega _\sigma
\omega _{\sigma '} }}\frac{{\epsilon _\sigma ^\dag  \frac{{\partial
D}} {{\partial k_x }}\epsilon _{\sigma '} \epsilon _{\sigma '}^\dag
\frac{{\partial D}} {{\partial k_y }}\epsilon _\sigma  }} {{(\omega
_\sigma   - \omega _{\sigma '} )^2 }}}.
\end{equation}
Because of Eq.~(\ref{eq-sew1}) and the following property:
\begin{equation}\label{eq-sDprpt}
  D_{ab}(-{\bf k})=D_{ab}^*({\bf k})= D_{ba}({\bf k}),
\end{equation}
we can transform from the positive-frequency bands to the negative-frequency band, and obtain
\begin{equation}
\kappa_{xy}^{(1)} = \frac{\hbar } {{8V  T}}\sum\limits_{{\bf
k},\sigma \sigma ' > 0} {[f(\omega _\sigma ) - f(\omega _{\sigma
'})](\omega _\sigma  + \omega _{\sigma '})^2 \frac{i} {{4 \omega
_\sigma  \omega _{\sigma '} }}\frac{{\epsilon _\sigma ^\dag
\frac{{\partial D}} {{\partial k_x }}\epsilon _{\sigma '} \epsilon
_{\sigma '}^\dag  \frac{{\partial D}} {{\partial k_y }}\epsilon
_\sigma  }} {{(\omega _\sigma   - \omega _{\sigma '} )^2 }}}.
\label{eq-skxy1}
\end{equation}
Here, $\sigma ,\sigma'$ can be both positive or negative.

Second, we calculate $\kappa_{ab}^{(2)}$. Utilizing the results
\begin{equation}
\begin{array}{ll}
\langle a_i^\dagger a_j^\dagger a_k a_l \rangle_{\rm eq} =
f_i f_j (\delta_{ik}\delta_{jl} + \delta_{il} \delta_{jk});\\
\langle a_i a_j a_k^\dagger a_l^\dagger \rangle_{\rm eq} =
(1+f_i)(1+ f_j) (\delta_{ik}\delta_{jl} + \delta_{il} \delta_{jk}),
\end{array}
\end{equation}
and the relation $f(-\omega)=-1-f(\omega)$, after some algebraic
derivation similar to the above, we obtain
\begin{equation}
\kappa_{xy}^{(2)} =
\frac{\hbar }
{{8V  T}}\sum\limits_{{\bf k},\sigma \sigma ' < 0} {[f(\omega _\sigma ) - f(\omega _{\sigma '})](\omega _\sigma  + \omega _{\sigma '})^2 \frac{i}
{{4 \omega _\sigma  \omega _{\sigma '} }}\frac{{\epsilon _\sigma ^\dag  \frac{{\partial D}}
{{\partial k_x }}\epsilon _{\sigma '} \epsilon _{\sigma '}^\dag  \frac{{\partial D}}
{{\partial k_y }}\epsilon _\sigma  }}
{{(\omega _\sigma   - \omega _{\sigma '} )^2 }}}.
\label{eq-skxy2}
\end{equation}
Therefore, the total phonon Hall conductivity can be written as
\begin{equation}
\kappa_{xy} =
\frac{\hbar }
{{8V  T}}\sum\limits_{{\bf k},\sigma\neq\sigma '} {[f(\omega _\sigma ) - f(\omega _{\sigma '})](\omega _\sigma  + \omega _{\sigma '})^2 \frac{i}
{{4 \omega _\sigma  \omega _{\sigma '} }}\frac{{\epsilon _\sigma ^\dag  \frac{{\partial D}}
{{\partial k_x }}\epsilon _{\sigma '} \epsilon _{\sigma '}^\dag  \frac{{\partial D}}
{{\partial k_y }}\epsilon _\sigma  }}
{{(\omega _\sigma   - \omega _{\sigma '} )^2 }}}.
\end{equation}
We can prove $\kappa_{xy}=-\kappa_{yx}$, such that
\begin{equation}
\kappa_{xy} =
\frac{\hbar }
{{16V  T}}\sum\limits_{{\bf k},\sigma\neq\sigma '} {[f(\omega _\sigma ) - f(\omega _{\sigma '})](\omega _\sigma  + \omega _{\sigma '})^2 \Omega_{k_x k_y}^{\sigma \sigma '}},
\end{equation}
\begin{equation}
\Omega_{k_x k_y}^{\sigma \sigma '}=\frac{i}
{{4 \omega _\sigma  \omega _{\sigma '} }}\frac{\epsilon _\sigma ^\dag  \frac{{\partial D}}
{{\partial k_x }}\epsilon _{\sigma '} \epsilon _{\sigma '}^\dag  \frac{{\partial D}}
{{\partial k_y }}\epsilon _\sigma - \epsilon _\sigma ^\dag  \frac{{\partial D}}
{{\partial k_y }}\epsilon _{\sigma '} \epsilon _{\sigma '}^\dag  \frac{{\partial D}}
{{\partial k_x }}\epsilon _\sigma  }
{(\omega _\sigma   - \omega _{\sigma '} )^2 }.
\label{eq-somega}
\end{equation}
Because of $\Omega_{k_x k_y}^{\sigma \sigma '}=-\Omega_{k_x
k_y}^{\sigma' \sigma}$, the phonon Hall conductivity can be written
eventually as
\begin{equation}\label{eq-skxy}
\kappa_{xy} =
\frac{\hbar }
{{8V  T}}\sum\limits_{{\bf k},\sigma\neq\sigma '} {f(\omega _\sigma ) (\omega _\sigma  + \omega _{\sigma '})^2 \Omega_{k_x k_y}^{\sigma \sigma '}},
\end{equation}
where $V$ is the total volume of $N=N_L^2$ unit cells. In the above
formula, the phonon branch $\sigma$ includes both positive and
negative values without restriction. We start with the positive
frequency bands to derive the conductivity formula. Through some
transformations, we finally obtain the simplified formula for phonon
Hall conductivity which combines the contribution from all the
frequency bands. The formula Eq.~(\ref{eq-skxy}) is different from
that given in Ref.~\cite{wang09}. In Ref.~\cite{wang09} the
contribution for phonon Hall conductivity from $J_2$ was omitted,
which is incorrect.

From the Eq.~(\ref{eq-sdisper}), we obtain
\begin{equation}
\epsilon_{{-\bf k},\sigma}^*(-A) = \epsilon_{{\bf k},\sigma}(A); \; \omega_{{-\bf k},\sigma} = \omega_{{\bf k},\sigma},
\label{eq-sew}
\end{equation}
and because of $D({\bf k})=D^*(-{\bf k}),\; \epsilon _\sigma ^T  \frac{{\partial D^*}}
{{\partial k_x }}\epsilon _{\sigma '}^* =\epsilon _{\sigma '} ^\dag  \frac{{\partial D}}
{{\partial k_x }}\epsilon _{\sigma}$, we have
\begin{equation}
  \Omega _{k_x k_y }^{\sigma \sigma '}({\bf k},-A)= \Omega _{k_y k_x }^{\sigma \sigma '}(-{\bf k},A)=-\Omega _{k_x k_y }^{\sigma \sigma '}(-{\bf k},A).
\end{equation}
So we obtain
\begin{equation}
    \kappa_{xy} (-A)=\kappa_{yx}(A)=- \kappa_{xy}(A).
\end{equation}
The Onsager reciprocal relations are satisfied.

If the system possesses the symmetry which satisfies
\begin{equation}
 SDS^{ - 1} = D,\; SAS^{ - 1}  = - A,
\end{equation}
where $S$ represents any symmetric operation, and from
Eq.~(\ref{eq-sdisper}), we obtain
\begin{equation}
 S \epsilon (A) = \epsilon (-A).
\end{equation}
Using the definition of the dynamic matrix $D=- A^2+\sum_{l'} K_{l,l'} e^{i({\bf R}_{l'} - {\bf
R}_{l})\cdot {\bf k}}$ and $SDS^{ - 1} = D$, we can obtain
\begin{equation}
S \frac {\partial D}{\partial k_\alpha }S^{ - 1} = \frac {\partial
D}{\partial k_\alpha }, \;\;\;\;\;\;\;\;(\alpha=x, y)
\end{equation}
Inserting $S^{-1}S=I$ into Eq.(\ref{eq-somega}), we obtain
\begin{equation}
  \Omega _{k_x k_y }^{\sigma \sigma '}(-A)= \Omega _{k_x k_y }^{\sigma \sigma '}(A).
\end{equation}
Then it is easy to obtain $ \kappa_{xy} (-A)=\kappa_{xy}(A)$, and
because of the Onsager relation, one can easily obtain that
\begin{equation}
 \kappa_{xy}=0,\;\; {\rm if}\; SDS^{ - 1} = D,\; SAS^{ - 1}  = - A.
\end{equation}

\begin{figure}[t]
\includegraphics[width=0.6\columnwidth]{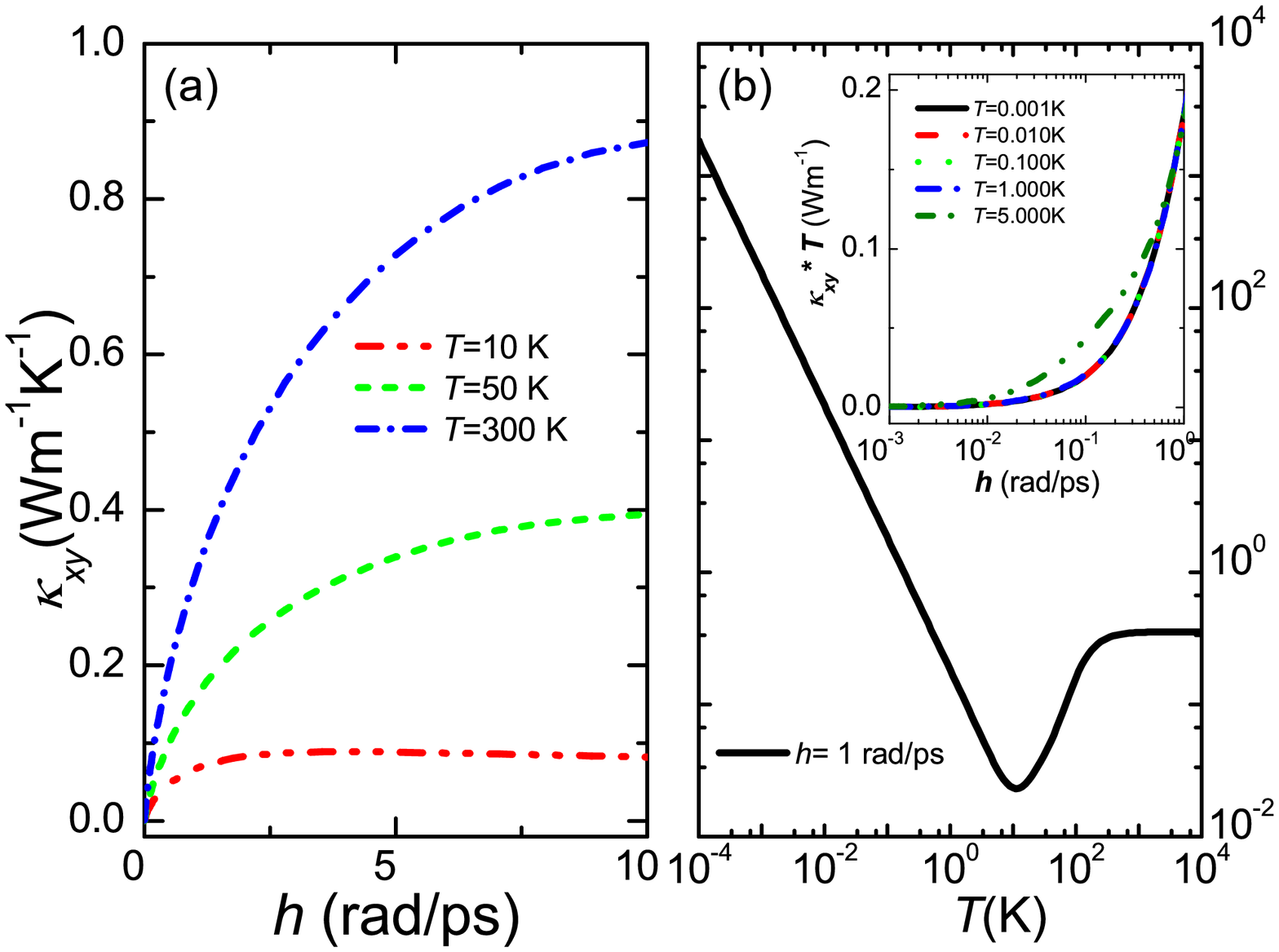}%
\caption{ \label{figS1kxyhT} (color online)  (a) Phonon Hall conductivity vs applied magnetic field for two-dimensional honeycomb lattice. (b) Phonon Hall conductivity vs temperature at fixed magnetic field $h=1$ rad/ps. The inset of (b) shows the product of phonon Hall conductivity and temperature $\kappa_{xy}T$ vs magnetic field $h$ for different temperatures.  }
\end{figure}

Fig.~\ref{figS1kxyhT}(a) shows the phonon Hall conductivity with
magnetic field for different temperatures.  In the weak magnetic
field range, the phonon Hall conductivity $k_{xy}$ is proportional
to the magnetic field, which is consistent with all the experimental
and theoretical results. We plot phonon Hall conductivity with a
large range of temperatures in Fig.~\ref{figS1kxyhT}(b).  At very
low temperatures, the phonon Hall conductivity is proportional to
$1/T$. $k_{xy}T$ will be constant for different temperatures lower
than $1 K$. This is due to the contribution from
$\kappa_{xy}^{(2)}$: if $T\rightarrow 0$, $1+f\rightarrow 1$, then
the conductivity linear with $1/T$ tends infinity. While the
longitudinal thermal conductivity $\kappa_{xx}$ is infinite for any
temperature \cite{wang09}, thus when $T \rightarrow 0$, the
transverse Hall conductivity, $\kappa_{xy} \rightarrow \infty$, has
the ballistic property similar to the longitudinal one. If
temperature is very high, all the modes contribute to the thermal transport, and $f\simeq
 k_B T/(\hbar\omega)$, then the phonon Hall conductivity becomes a
constant, which can be seen in Fig.~\ref{figS1kxyhT}(b).

\subsection{\label{secberrycurv} THE BERRY PHASE AND BERRY CURVATURE}
Using the similar method proposed by Berry\cite{berry84}, we derive
the Berry phase and Berry curvature in the following. Starting from
\begin{equation}
i \frac{\partial } {{\partial t}}x(t) = H_{\rm eff} x(t)
\end{equation}
and substituting
$$
x(t) = e^{i\gamma _\sigma  (t) - i\int_0^t {dt'\omega _\sigma({\bf
k}(t'))}} x _\sigma  ({\bf k}(t)),
$$
we can obtain the Berry phase across the Brillouin zone as
$$
\gamma _\sigma   = \oint\limits{
 {\bf A}^\sigma ({\bf k}) d {\bf k} },  \;\;\;\;\;\;\;\;
 {\bf A}^\sigma ({\bf k}) = i\tilde x_\sigma ^T \frac{\partial } {{\partial {\bf
k}}}x _\sigma  .
$$
Here $ x_\sigma, \tilde x _\sigma^T$ correspond to the right and
left eigenvectors, and $ \tilde x _\sigma^T x _{\sigma '}  = \delta
_{\sigma \sigma '}$, $\sum\limits_\sigma  {x _\sigma  \tilde x
_\sigma ^T  = I}$. ${\bf A}^\sigma  ({\bf k})$ is the so-called
Berry vector potential.
Therefore the Berry curvature is obtained through the Stokes theorem
as:
\begin{equation}
\Omega _{k_x k_y }^\sigma   = \frac{\partial }
{{\partial k_x }}{\bf A}_{k_y }^\sigma   - \frac{\partial }
{{\partial k_y }}{\bf A}_{k_x }^\sigma   = i\sum\limits_{\sigma ' \ne \sigma } {\frac{{\tilde x _\sigma ^T \frac{{\partial H_{\rm eff} }}
{{\partial k_x }}x _{\sigma '} \tilde x _{\sigma '}^T \frac{{\partial H_{ \rm eff} }}
{{\partial k_y }}x _\sigma   - (k_x  \leftrightarrow k_y )}}
{{(\omega _\sigma   - \omega _{\sigma '} )^2 }}}
\end{equation}
Inserting the vector $x$ and the expression of matrix $H_{\rm eff}$,
we obtain
\begin{equation}\label{eq-sbrycv}
\Omega _{k_x k_y }^\sigma
 = \sum\limits_{\sigma ' \ne \sigma } {\frac{i}
{{4\omega _\sigma  \omega _{\sigma '} }}\frac{{\epsilon _\sigma ^\dag  \frac{{\partial D}}
{{\partial k_x }}\epsilon _{\sigma '} \epsilon _{\sigma '}^\dag  \frac{{\partial D}}
{{\partial k_y }}\epsilon _\sigma   - (k_x  \leftrightarrow k_y )}}
{{(\omega _\sigma   - \omega _{\sigma '} )^2 }}}= \sum\limits_{\sigma ' \ne \sigma } {\Omega _{k_x k_y }^{\sigma \sigma '} }
\end{equation}
where $\Omega _{k_x k_y }^{\sigma \sigma '}$ indicates the
contribution to the Berry curvature of the band $\sigma$ from a different
band $\sigma '$. Therefore, the phonon Hall conductivity formula
Eq.~(\ref{eq-skxy}) can be interpreted in terms of the Berry
curvature.

\subsection{\label{secchern} THE CALCULATION OF THE CHERN NUMBER}
The topological Chern number is obtained by integrating the Berry
curvature over the first Brillouin zone as
\begin{equation}\label{eq-scherni}
C^\sigma   = \frac{1} {{2\pi }}\int_{\bf BZ} {dk_x dk_y \Omega _{k_x k_y
}^\sigma  }.
\end{equation}
For numerical calculation, we use
\begin{equation}\label{eq-schernn}
C^\sigma   = \frac{{2\pi }} {{L^2}}\sum\limits_{\bf k} {\Omega _{k_x
k_y }^\sigma  }.
\end{equation}
where ${\frac{1} {L^2}\sum\limits_{\bf k}= \int {\frac{{dk_x dk_y }}
{{(2\pi )^2 }}}}$ and $V=L^2a$, $L^2$ is the area of the sample.
\begin{figure}[t]
\includegraphics[width=0.6\columnwidth]{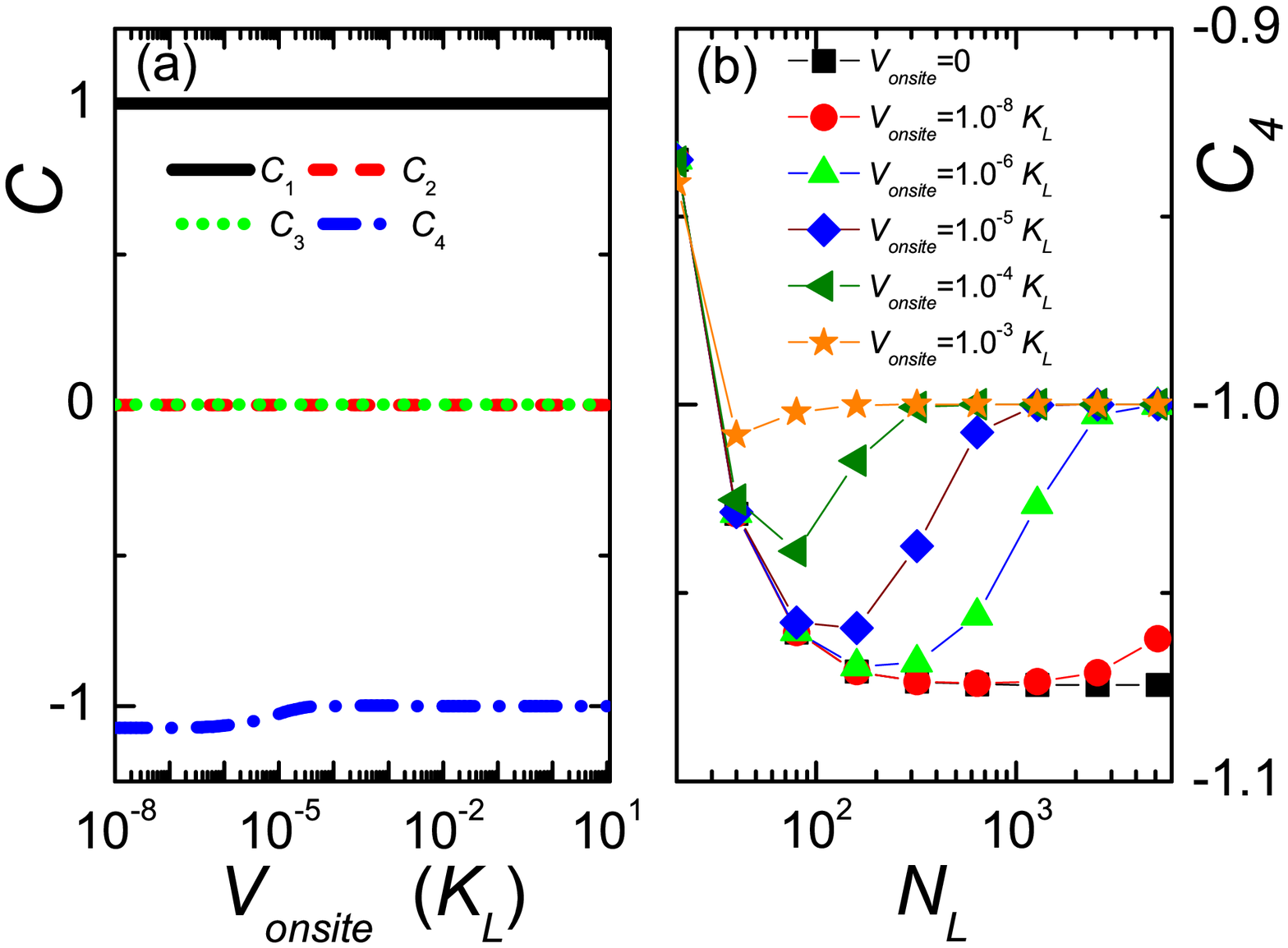}%
\caption{ \label{figS2chern} (color online) (a) The four Chern
number vs onsite potential $V_{\mathrm{onsite}}$. The unit for
onsite potential is longitudinal spring constant $K_L$. Here
$N=N_L^2=160000$;  (b) The Chern number of the fourth band
changes with $N_L$ for different onsite potentials. For both (a) and
(b), $h=1$ rad/ps. }
\end{figure}

To calculate the integer Chern numbers, large ${\bf k}$-sampling
points $N$ is needed. However there is always a zero eigenvalue at
the $\Gamma$ point of the dispersion relation, which corresponds to
a singularity of the Berry curvature. Therefore, we cannot sum up
the Berry curvature very near this point to obtain Chern number of
this band, unless we add a negligible on-site potential
$\frac{1}{2}u^TV_{\mathrm{onsite}}u$ to the original Hamiltonian. In
Fig.~\ref{figS2chern}(a), without the on-site potential, the Chern
number of the fourth band is not an integer, no matter how large the
sample size $N=N_L^2$ is (see Fig.~\ref{figS2chern}(b)). If we add
the external on-site potential, the Chern number of the fourth band
will become integer. In Fig.~\ref{figS2chern}(a), the $C_4$ changes
gradually to $-1$ with increasing the on-site potential, while other
Chern numbers do not change. And from Fig.~\ref{figS2chern}(b), we
see that with larger on-site potential, the Chern number of the
fourth band could be an integer for smaller sample sizes.
\end{widetext}

\end{document}